\RequirePackage{fix-cm}
\documentclass[smallextended]{svjour3}       
\smartqed  
\usepackage{graphicx}
\usepackage{epsfig}
\usepackage{amsmath}
\usepackage{soul}
\usepackage{color}
\usepackage{bm}
\usepackage{epsfig}
\usepackage{natbib}
\usepackage{multirow}
\usepackage{rotating}
\usepackage{booktabs}
\usepackage{hyperref}

\newcommand{\tr}{^{\prime}}
\def\b#1{\mbox{\boldmath $#1$}}    
\def\cg#1{\mbox{${\cal #1}$}}      

\renewcommand{\th}{\theta}

\newcommand{\be}{\beta}

\newcommand{\ga}{\gamma}

\begin{document}

\title{A multilevel finite mixture item response model 
to cluster examinees and schools 
}

\titlerunning{Multilevel finite mixture IRT}        
\author{     Michela Gnaldi   \and
       Silvia Bacci  \and
        Francesco Bartolucci
}




\maketitle
\begin{abstract}
Within the educational context, a key goal is to assess students' acquired skills and to cluster students according to their ability level. 
In this regard, a relevant element to be accounted for is the possible effect of the school students come from.
For this aim, we provide a methodological tool which takes into account the multilevel structure of the data (i.e., students in schools) in a suitable way.
This approach allows us to cluster both students and schools into homogeneous classes of ability and effectivness, and to assess the effect of certain students' and school characteristics on the
probability to belong to such classes.  
The approach relies on an extended class of multidimensional latent class IRT models characterized by: ({\em i}) latent traits defined at student level and at school level, ({\em ii}) latent traits represented through random vectors with a discrete distribution, ({\em iii}) the inclusion of covariates at student level and at school level, and ({\em iv}) a two-parameter logistic parametrization for the conditional probability of a correct response given the ability. 
The approach is applied for the analysis of data collected by two national tests administered
in Italy to middle school students in June 2009: the INVALSI Italian Test and Mathematics Test. 
Results allow us to study the relationships between observed characteristics and latent trait standing within each latent class at the different levels of the hierarchy. They show that examinees' and school expected observed scores, at a given latent trait level, are dependent on both unobserved (latent class) group membership and observed first and second level covariates.
\end{abstract}

\section{Introduction}




In accordance with the assumption of unidimensionality, which characterises Classical Item Response Theory (IRT) models, responses to a set of items only depend on a single latent trait which, in the educational setting, can be interpreted as the studentÕs ability. If unidimensionality is not met, summarizing studentsÕ performances through a single score, on the basis of a unidimensional IRT model, may be misleading as test items indeed measure more than one ability.

To take the presence of classes of individuals into account in a suitable way, it is possible to rely on Latent Class (LC) analysis  \citep{laza:henr:68,good:74}, which  is based on the idea that the population under study is formed by a finite number of latent (i.e., unobservable) classes of subjects with homogeneous ability levels.  
More precisely, mixture IRT models are a combination of IRT models and LC analysis. 
The basic idea behind them is that the same IRT model holds within subjects in the same class, whereas  different IRT models hold among different classes.  
Thus, mixture IRT models allow for the discreteness of the ability distribution, which is made up of as support points as latent classes. 

An example we refer to in the following,
is the class of LC IRT models for binary items and ordinal polytomous items  \citep{bart:07, bart:bacci:gna}. These types of models assume fixed (rather than class-specific) item parameters among latent classes. Moreover, they take into account multidimensional latent traits \citep{rec:09} and more general item parameterizations than 
those of Rasch-type models  \citep{rasch:61}, such as the two-parameter logistic (2PL) 
model introduced by \cite{birn:68}. An approach related to that of \cite{bart:07} is due to \cite{dav:08}, who proposed a diagnostic model based on fixed rather than free abilities. For further examples of LC 
versions of IRT models we also recall \cite{lind:91}, \cite{for:92}, \cite{form:95}, \cite{hoi:mol:97}, \cite{ver:01}, and \cite{smit:03}.
See also \cite{mas:85}, \cite{lan:rost:88},  \cite{hei:96}, \cite{chr:02}, and  \cite{for:07}, 
which outline the greater flexibility of IRT models based on the discreteness assumption of the latent traits with respect to those based on the continuity assumption. 
Among others, one of the most known examples of discretized variants of IRT models is given by the mixed Rasch model for binary and ordinal polytomous data \citep{rost:90, rost:91, dav:rost:95}, built as a mixture of latent classes with a separate Rasch model, characterized by class-specific person parameters and class-specific item parameters, assumed to hold within each of these classes. 
Besides, from an applicative point of view, mixture IRT models have been used to set proficiency standards \citep{jia:all:12},
identify solution strategies  \citep {mis:ver:90}, study the effects of test speediness \citep {bol:coh:wol:02}, and to identify latent classes which differ with regards to the use of response scales \citep[see, for example,][]{mai:kel:fli:08}.

Another relevant element, which is ignored by traditional IRT models, concerns that part of unobserved heterogeneity of item responses due to  multilevel  structures of data, where individual units  are nested in groups, such as students within schools. In such a type of data, it is reasonable to assume that  students sharing a same school context are more similar than their colleagues belonging to different schools in terms of latent traits and, as a consequence, the corresponding item responses cannot in general assumed to be independent.  A wide class of models which accounts for the group effect and also  encloses the above discussed assumption of discreteness and multidimensionality of the latent traits is represented by the multilevel finite mixture models  \citep{ver:03, skr:rabe:04,  ver:08, gril:ramp:07,cho:coh:10, bart:pen:vit:11}; 
see also \cite{kamata:01}, \cite{maier:01}, and \cite{fox:05} for multilevel IRT models under the assumption of normality of latent traits. 
Overall, this type of models is based on specifying different latent traits at each hierarchical level. Latent traits at individual level are measured through the test items, as usual in the IRT setting, but their level is affected by the hierarchical upper level units.
More precisely, at the group level, one can specify one or more latent traits that represent such a unobserved group effect. 
Besides, as a consequence of  the discreteness assumption, a multilevel finite mixture model allows us to cluster both individuals and groups in classes, which are homogeneous with respect to the corresponding latent traits. 

A further element which is be of interest, other than the measurement of the latent traits, is represented by the possible effect of one or more 
observed characteristics on the level of the latent traits.
In principle, a simple procedure to account for such covariate effects involves three consecutive steps: ({\em i}) building a mixture IRT model for a set of response variables, ({\em ii}) assigning subjects to the latent classes based on their posterior probabilities, and ({\em iii}) evaluating the association between class membership and external variables of interest (i.e., gender, age, etc.) using cross-tabulations or multinomial logistic regression analysis \citep {ver:10}.
However, it has been demonstrated \citep {bol:cro:hag:04} that this procedure substantially underestimates the association between class membership and covariates. Subsequent studies introduced correction methods that involve modifying the third step \citep {ver:10}. 
Moreover,  a number of studies \citep[see, for instance,][]{smi:kel:fli:99,smi:kel:fli:00} showed that by incorporating collateral information into various mixture IRT models  reduces the standard errors in the item parameter estimates considerably. The same authors also demonstrated through simulation studies that latent class assignment can benefit substantially from incorporating external variables that associate with the latent classes, especially when the sample size is large.
 Most recently, 
within a Differential Item Functioning (DIF) framework, further studies \citep[see, for instance,][]{tay:new:ver:11,tay:ver:wan:13} proposed to overcome such a three step procedure through an Item Response Theory with Covariates (IRT-C) procedure which allows us 
to evaluate the effect of class membership and external variables simultaneously.   
In multilevel settings, covariates may express examinees' characteristics (e.g., gender) as well as schools' characteristics (e.g., geographic area) and one can speak about level 1 and level 2 covariates, respectively. 
Level 1 covariates contribute, with level 2 latent traits,  to explain students' level latent traits, whereas level 2 covariates contribute to explain school level latent traits. 

In line with the above issues, we illustrate a multilevel extension of the class of LC IRT models developed by \cite{bart:07} to include the group effect due to the aggregation of examinees in different schools, other than  the effects of covariates.
Specifically, we assume the presence of a single latent trait at the school level which affects the abilities at the students' level. Such individual abilities are then measured through a multidimensional version of LC IRT models based on a 2PL parameterisation for the conditional probability of a given response, in the case of binary items. 
The proposed model is estimated through the maximum marginal likelihood method making use of the Expectation-Maximization (EM) algorithm \cite{demp:lair:rubi:77}, avoiding in this way the use of three steps methods.
The described estimation method is implemented in the {\tt R} package {\tt MultiLCIRT}, freely downloadable from \verb'http://www.CRAN.R-project.org/package=MultiLCIRT'.;
see \cite{bart:bacci:gna} for a detailed description of this package in the case of no covariates affecting the latent ability level.

The proposed multilevel LC IRT model with covariates is illustrated through the analysis of the abilities measured by two Italian National Tests for the assessment of primary, lower middle,
and high-school students, which are developed and yearly collected by the
National Institute for the Evaluation of the Education System
(INVALSI).
Here we focus on the INVALSI Tests administered to middle school
students as they are having an increasing relevance in the Italian
education context and their collection will become compulsory in the
near future. These data are based on
a nationally representative sample of 27,592 students within 1,305
schools and refer to students' performances in the Italian Test and the
Mathematics Test, administered in June 2009. 

With reference to the data mentioned above, we are interested in detecting examinees 
and schools heterogeneity, studying the latent score distribution and the size of the latent classes, and  examining the relationship between observed covariates and latent trait standing within each latent class, at both hierarchical levels. 
The key aim of this work is twofold: ({\em i}) clustering students and schools into homogeneous classes of latent traits, evaluating, on one side, the degree to which latent subgroups of examinees show distinct response strategies and, on the other side, the degree to which 
latent subgroups (or types) of schools differently characterize the expected abilities of their students, and ({\em ii}) assessing if and how examinees' and school covariates (i.e., gender and geographic area, respectively) affect the probability for a student or for a school to belong to each of these latent classes.

The remainder of this paper is organized as follows. In the next section we
describe the INVALSI data used in our analysis. The statistical methodological
approach employed to investigate the structure of the questionnaires is described
in Section \ref{sec:methods}. First, we recall the basics for the model adopted
in our study; then, we illustrate the extension to take into account the multilevel structure of the data and the covariate effects.
Details about the estimation algorithm and the use of these models are given in Section \ref{sec:ML}.
Finally, in Section \ref{sec:appl}, we illustrate the main results obtained by
applying the proposed approach to the INVALSI datasets and in Section \ref{sec:conc} we draw the main conclusions of the study.
\section{The INVALSI data}
The INVALSI Italian Test and Mathematics Test were administered in June 2009, at the end of the
pupils' compulsory educational period. Afterwards, a nationally
representative sample made of 27,592 students was drawn \citep{INV:093}.
From each of the 21
strata, corresponding to the 21 Italian geographic regions, a sample of schools was
drawn independently; allocation of sample units within each
stratum was chosen proportional to an indicator based on the
standard deviations of certain variables 
(e.g., school size)
and the stratum sizes.
Classes within schools were then sampled through a random procedure, with one class sampled in each school.
Overall, 1305 schools (and classes) were sampled. 

The INVALSI Italian Test includes two sections, a Reading Comprehension
section and a Grammar section. The first section is based on
two texts: a narrative type text (where readers engage with imagined events and
actions) and an informational text (where readers engage with real settings);
see \cite{INV:091}.
The comprehension processes are measured by 30 items, which require
students to demonstrate a range of abilities and skills in constructing meaning
from the two written texts. Two main types of comprehension processes were
considered in developing the items: Lexical Competency, which covers the ability to make
sense of worlds in the text and to recognize meaning connections among them, and
Textual
Competency, which relates to the ability to: ({\em i}) retrieve or locate
information in
the text, ({\em ii}) make inferences, connecting two or more ideas or pieces of
information and recognizing their relationship, and ({\em iii}) interpret and
integrate
ideas and information, focusing on local or global meanings. The Grammar section
is made of 10 items, which measure the ability
of understanding the morphological
and syntactic structure of sentences within a text.

The INVALSI Mathematics Test consists of 27 items covering four main content
domains: Numbers, Shapes and Figures, Algebra, and Data and Previsions
\citep{INV:092}. The Number content domain consists of understanding (and
operation with) whole numbers, fractions and decimals, proportions, and percentage
values. The Algebra domain requires students the ability to understand, among
others, patterns, expressions and first order equations, and to represent them
through words, tables, and graphs. Shapes and Figures covers topics such as
geometric shapes, measurement, location, and movement. It entails the ability to
understand coordinate representations, to use spatial visualization skills in
order to move between two and three dimensional shapes, draw symmetrical figures,
and understand and being able to describe rotations, translations, and reflections
in mathematical terms. The Data and Previsions domain includes three main topic
areas: data organization and representation (e.g., read, organize and display data
using tables and graphs), data interpretation (e.g., identify, calculate and
compare characteristics of datasets, including mean, median, mode), and chance
(e.g., judge the chance of an outcome, use data to estimate the chance of future
outcomes).

All items included in the Italian Test are
of multiple choice type, with one
correct answer and three distractors, and are dichotomously scored
(assigning 1 point to  correct answers and 0 otherwise).
The Mathematics Test is also made
of multiple choice items, but it also contains two open questions for which
a score of 1 was assigned to correct answers and  a score of 0 to incorrect and partially correct answers.

Preliminary analyses (see Table~\ref{table1})
show that students' performances on Test items were different on
account of students' gender (male (M),  female (F)) and school geographic area (North-West
 (NW), North-East (NE), Centre, South, and Islands). Overall, females
performed better than males in the Italian Test, but worse than
males in the Mathematics Test. In both Tests, average percentage
scores per geographic area revealed very different 
levels of
attainment. 

\begin{table}[ht]
\caption{Average relative score per gender and geographic area for the three dimensions of the 
INVALSI Tests.}\label{table1}
\begin{center}
\begin{tabular}{llcccccc}
  \hline\hline
  && \multicolumn5c{area}\\\cline{3-7}
 & & NW  &   NE  &   Centre  &   South   &   Islands & Total \\ 
  & &  (17.70\%)   &    (20.95\%)  &   (20.39\%)  &   (20.91\%)   &    (20.05\%) &  \\ 
  M    (50.28\%)   & V1 & 0.730 & 0.711& 0.731 & 0.725 & 0.701 &0.721 \\ 

 & V2& 0.781 & 0.743 & 0.780 & 0.815 & 0.781 & 0.780 \\ 
      & V3& 0.770 & 0.760 & 0.783 & 0.795 & 0.789 & 0.779\\\hline 
  F    (49.72\%)  & V1 & 0.744 & 0.737 & 0.755 & 0.744 & 0.732 & 0.743 \\ 
  & V2 & 0.810 & 0.778 & 0.816 &  0.836 & 0.812 &0.811\\ 
      & V3 & 0.744 & 0.732 & 0.769 & 0.792 & 0.781 &  0.764   \\\hline 
    Total &  &  0.763 & 0.743  &  0.772 & 0.785 & 0.767  &   0.766  \\  				
   \hline
\end{tabular}
\end{center}
\end{table}



%
\section{Methodological approach}\label{sec:methods}
%
In this section, we illustrate  the methodological approach adopted
to investigate the students' abilities. First, we review the basic
model proposed by  \cite{bart:07} and then we extend it to 
the multilevel setting.

\subsection{Preliminaries}\label{sec:preliminaries}

The class of multidimensional LC IRT models developed by \cite{bart:07} presents two main differences with respect to classic IRT models: (\textit{i}) the latent structure is multidimensional and (\textit{ii}) it is based on latent variables that have a discrete distribution.
We consider in particular the version of these models based on the two-parameter (2PL) logistic
parameterisation of the conditional response probabilities \citep{birn:68}.

Let $n$ denote the number of subjects in the sample and suppose that
these subjects answer $r$ dichotomous test items 
\citep[see][for a more general formulation for polytomously-scored items]{bacci:bartolucci:gnaldi:2014} which measure $s$ different latent traits or dimensions.
For the moment the possibility is ignored that these $n$ subjects might be nested in different observed groups. %
 Also let $\mathcal{J}_d$, $d
= 1,\ldots,s$, be the subset of $\mathcal{J}=\{1,\ldots,r\}$
containing the indices of the items measuring the latent trait of
type $d$ 
and let $r_d$ denoting the cardinality of this subset, so
that $r=\sum_{d=1}^s r_d$. Since we assume that each item measures
only one latent trait, the subsets $\mathcal{J}_d$ are disjoint;
obviously, these latent traits may be correlated. Moreover, adopting
a 2PL parameterisation, it is assumed that
%
\begin{equation}
\textrm{logit}[p(Y_{ij}=1\mid V_i = v_i)] =
\gamma_j\left(\sum_{d=1}^{s} \delta_{jd}\xi_{v_id}^{(V)} - \beta_j\right),
\quad i=1,\ldots,n,\:j = 1,\ldots,r. \label{eq:multid2PL}
\end{equation}
In the above expression, $Y_{ij}$ is the random variable
corresponding to the response to item $j$ provided by subject $i$
($Y_{ij}=0,1$ for wrong or right response, respectively). Moreover,
$\beta_j$ is the difficulty level of item $j$ and $\gamma_j$ is its discriminating level, 
$V_i$ is a latent variable indicating the latent class of the subject, 
$v_i$ denotes one of the possible realizations of $V_i$, and $\delta_{jd}$ is a dummy variable
equal to $1$ if 
index $j$ belongs to $\mathcal{J}_d$ (and then 
item $j$
measures the $d$th latent trait) and to 0 otherwise. Finally, a
crucial assumption is that each random variable $V_i$ has a
discrete distribution with support $\{\b\xi_1^{(V)},\ldots,\b\xi_{k_V}^{(V)}\}$,
which correspond to $k_V$ 
latent classes in the population. 
Associated to subjects in latent class $v$ there is a vector $\b\xi_v^{(V)}$ with elements $\xi_{vd}$ corresponding to the ability level of subjects in latent class $v$ with respect to dimension $d$.
Note that, when $\ga_j=1$ for all $j$, then the above 2PL
parameterisation reduces to a multidimensional Rasch
parameterisation. At the same time, when the
elements of each support vector $\b\xi_v^{(V)}$ are obtained by the same
linear transformation of the first element, the model is indeed
unidimensional even when $s>1$. 

As for the conventional LC model \citep{laza:henr:68, good:74}, the
assumption that the latent variables have a discrete distribution
implies the following {\em manifest distribution} of the full
response vector $\b Y_i= (Y_{i1},\ldots,Y_{ir})\tr$:
\begin{equation}
p(\b y_i)=p(\b Y_i=\b y_i) = \sum_{v=1}^k p_v(\b y_i)\pi_{v}^{(V)},\label{equ:1}
\end{equation}
where $\b y_i=(y_{i1},\ldots,y_{ir})\tr$ denotes a realization of $\b Y_i$,
$\pi_{v}^{(V)} = p(V_i=v)$ is the weight of the $v$th latent 
class, with $\sum_{v}\pi_{v}^{(V)} = 1$ and $\pi_{v}^{(V)}>0$ for $v=1,\ldots,k_V$, and, due to the {\em local independence assumption} which characterises all IRT models, we have
\begin{equation}
p_v(\b y_i)=
p(\b Y_i=\b y_i\mid V_i=v)=
\prod_{j=1}^r p(Y_{ij}=y_{ij} \mid V_i=v), \quad v=1,\ldots,k_V.
\label{eq:cond_prob}
\end{equation}

The specification of the multidimensional LC 2PL model, based on the
assumptions illustrated above, univocally depends on: ({\em i}) the
number of latent classes ($k_B$), ({\em ii}) the number of the
dimensions ($s$), and ({\em iii}) the way items are associated to
the different dimensions. The last feature is related to the
definition of the subsets $\mathcal{J}_d$, $d = 1,\ldots,s$. 

%

%
\subsection{Extension to multilevel setting}\label{sec:latregr}
%
The model illustrated above does not take into account the dependence between item responses of individuals belonging to a same group, which typically rises in presence of multilevel data, as well as the possible effect of one or more observed covariates. 
More in general, it is reasonable to suppose that the probability of a subject coming from a given latent class of ability is influenced  by some unobserved characteristics of the group which she/he belongs to.
Such unobserved characteristics define a new latent trait that can be called ``group effect'', which add up to the effect of individual covariates and which may be explained by some group covariates.  
 
Let  $Y_{hij}$ denote   the response provided by subject $i$ within group $h$ to item $j$, with possible values 0 and 1.
Note that such notation 
is something different by that usually adopted in the multilevel models setting \citep{gold:11}, where the first subscript denotes the individual units and the second subscript denotes the group units.
However, the  reversed order of subscripts allows us to accommodate the third subscript $j$, which refers to the items, coherently with notation typically adopted in the IRT literature.
 
Let $n_h$ be  the size of group $h$,  $H$ the number of groups, and  $\b W_h = (W_{h1}, \ldots, W_{hm_U})\tr$ a vector of $m_U$ covariates (group level covariates) related to this vector.
Similarly, let $\b X_{hi} = (X_{hi1}, \ldots, X_{him_V})\tr$ denote a vector of $m_V$ covariates (individual level covariates) for subject $i$ in group $h$.
Besides, according to the definitions given in the previous section, the distribution of the latent traits measured by the questionnaire is described by a latent variable $V_{hi}$ with $k_V$ support points, whereas the group effect is denoted  by a discrete latent variable $U_h$ with $k_U$ support points, $\{\xi_1^{(U)},\ldots,\xi_{k_U}^{(U)}\}$.
The $k_V$ and $k_U$ support points define as many latent classes of individuals and groups, respectively.
To avoid any misunderstanding, we use hereafter the term ``type'', as a synonymous of latent class, when we refer to the latent trait distribution at the group level, that is, $U_h$. 
 
The relation between $V_{hi}$ and $Y_{hij}$ is based on the formulation illustrated in Section \ref{sec:preliminaries}, equation (\ref{eq:multid2PL}), where subscript $h$ is added to account for the aggregation of subjects in groups:
%
\begin{equation}
\textrm{logit}[p(Y_{hij}=1\mid V_{hi} = v_{hi})] =
\gamma_j\left(\sum_{d=1}^{s} \delta_{jd} v_{hid} - \beta_j\right),
\quad i=1,\ldots,n, \:h = 1, \ldots, H,\:j = 1,\ldots,r. 
\label{eq:multid2PL_bis}
\end{equation}
Moreover, as now $V_{hi}$ depends on $U_h$ and  $\b X_{hi}$, then in equation (\ref{equ:1})
 there is not any more a constant
weight $\pi_{v}^{(V)} = p(V_{hi}=v)$ 
for each latent class, but a weight $\pi_{hi,v|u}^{(V)} = p(V_{hi}=v|U_h=u,\b X_{hi}=\b x_{hi})$  
depending on the value $\xi_u^{(U)}$ of latent trait $U_h$ and on the individual covariate configuration $\b X_{hi}=\b x$. 

The above dependence is represented by a multinomial logit parameterisation 
\citep{dayton:88, for:07b} for weights $\pi_{hi,v|u}^{(V)}$, $v=2,\ldots,k_V$,
with respect to $\pi_{hi,1|u}$ (or another weight), as follows
\begin{equation}
\log\frac{\pi_{hi,v|u}^{(V)}}{\pi_{hi,1|u}^{(V)}}=
\zeta_{0uv}^{(V)}+\b x_{hi}\tr\b\zeta_{1uv}^{(V)},\quad v=2,\ldots,k_V,
\label{eq:multi1}
\end{equation}
where elements of vector $\b\zeta_{1uv}^{(V)}$ denote the effect of  individual covariates  $\b X_{hi}$ on the logit of $\pi_{hi,v|u}^{(V)}$ with respect to $\pi_{hi,1|u}^{(V)}$ and $\zeta_{0uv}^{(V)}$ is the intercept specific for examinees of class $v$  that belong to  a school in class (or type) $u$.

Let  $\pi_{hu}^{(U)}=p(U_h=u|\b W_h=\b w_h)$ denote the weights associated to support points of $U_h$ distribution that depend on the group covariate configuration $\b W_h=\b w_h$. Then, a similar multinomial logit parameterisation is  adopted as concerns the conditional distribution of $U_h$ given $\b W_h$:
\begin{equation}
\log\frac{\pi_{hu}^{(U)}}{\pi_{h1}^{(U)}}=
\zeta_{0u}^{(U)}+\b w_h\tr\b\zeta_{1u}^{(U)},\quad u=2,\ldots,k_U,
\label{eq:multi2}
\end{equation}
where  elements of vector $\b\zeta_{1u}^{(U)}$ denote the effect of  group covariates  $\b W_{h}$ on the logit of $\pi_{hu}^{(U)}$ with respect to $\pi_{h1}^{(U)}$ and $\zeta_{0u}^{(U)}$ is the intercept specific for schools within type $u$.



To conclude, it has to be outlined that the multilevel multidimensional LC IRT model may be equivalently depicted both as a two-level random intercept logit model for multivariate responses and as a three-level random intercept logit model for a univariate response \citep{ver:08}. In the first case, item responses are considered as a multivariate dependent variable and the hierarchical structure is represented by individuals nested within groups. In the second case,  item responses represent a further hierarchical level (item responses within individuals within groups). In both cases, latent traits involved at individual and group levels represent the random intercepts.

\section{Likelihood based inference}\label{sec:ML}
In this section, we deal with maximum likelihood of the extended model based on assumptions (\ref{eq:multi1}) and (\ref{eq:multi2}) and the problem of selecting the number of latent classes and testing hypotheses on the parameters. 
The hypotheses of greatest interest in our context are those of absence of effect of the covariates on the latent distributions at cluster and individual levels.
\subsection{Maximum likelihood estimation}
%
For  given $k_U$ and $k_V$, the parameters of the proposed model may be estimated by maximizing the log-likelihood
\begin{eqnarray}
\ell(\b\th)&=&\sum_{h=1}^H\log\sum_{u=1}^{k_U}\pi_{hu}^{(U)}\rho_h(\xi_u^{(U)}),\label{eq:loglik2}
\end{eqnarray}
where $\b\th$ is the vector containing all the free parameters,
and 
\begin{eqnarray*}
\rho_h(\xi_u^{(U)})&= &\prod_{i=1}^{n_h}\sum_{v=1}^{k_V}\pi_{hi,v|u}^{(V)}
\prod_{j=1}^r p(y_{hij}|V_{hi}=v),
\end{eqnarray*}
with $p(y_{hij}|V_{hi}=v)$ defined as in (\ref{eq:multid2PL_bis}).

About the vector $\b\th$, we clarify that it contains  item parameters
$\be_j$ (difficulty) and $\ga_j$ (discriminating index),   covariate parameters $\b\zeta_{1uv}^{(V)}$ and $\b\zeta_{1u}^{(U)}$,
ability levels  $\b\xi_{v}^{(V)}$ and $\xi_u^{(U)}$ affecting the 
individual and group weights $\pi_{hi,v|u}^{(V)}$ and $\pi_{hu}^{(U)}$. However, to make  the model identifiable, we adopt
the constraints
\[
\beta_{j_d}=0,\:\gamma_{j_d}=1,\quad d=1,\ldots,s,
\]
with $j_d$ denoting a reference item for the $d$-th dimension
(usually, but not necessarily, the first item for each latent trait). 
In this way, for each item $j$, with $j\in\cg J_d\setminus\{j_d\}$,
the parameter $\be_j$ is interpreted in terms of differential difficulty level of
this item with respect to item $j_d$; similarly, $\ga_j$, is
interpreted in terms of ratio between the discriminant index of item
$j$ and that of item $j_d$. 

Considering the above identifiability constraints,  the number of free parameters
collected in $\b\th$ is equal to 
\begin{equation*}\label{eq:numpar2}
\# \textrm{par} =    (k_V-1) (m_V + k_U) + (k_U-1) (m_U+1) + k_V s + 2(r-s)
\end{equation*} 
In fact, there are $(k_V-1) (m_V + k_U) + (k_U-1) (m_U+1)$ regression coefficients for the latent classes, $k_V s$  
ability parameters, $r-s$ discriminating parameters, and $r-s$ difficulty parameters. Under the Rasch parameterisation, the number of parameters decreases by $r-s$ as the discriminating indices have not to be estimated. Note that only the ability parameters $\b \xi_{v}^{(V)}$ are estimated, whereas  parameters $\xi_{u}^{(U)}$ are obtained as  average of $\b \xi_{v}^{(V)}$ values  weighted by $\pi_{hu}^{(U)}$.

In order to maximize the log-likelihood $\ell(\b\th)$, we make use of the Expectation-Maximization (EM) algorithm \citep{demp:lair:rubi:77}, which is developed along the same lines as in \cite{bart:07}; see also \cite{bart:pen:vit:11}. 

The {\em complete log-likelihood}, on which the EM algorithm is based,
may be expressed as
\begin{equation}\label{eq:comploglik}
\ell^*(\b\th)=\sum_{h=1}^H[\ell_{1h}^*(\b\th) +\ell_{2h}^*(\b\th)+\ell_{3h}^*(\b\th)],
\end{equation}
with
\begin{eqnarray*}
\ell_1^*(\b\th)&=&\sum_{i=1}^{n_h}\sum_{j=1}^J\sum_{v=1}^{k_V}z_{hiv}\log p(y_{hij}|V_{hi}=v),\\
\ell_2^*(\b\th)&=&\sum_{i=1}^{n_h}\sum_{u=1}^{k_U}\sum_{v=1}^{k_V}z_{hu}z_{hiv}\log \pi_{hi,v|u}^{(V)},\\
\ell_3^*(\b\th)&=&\sum_{u=1}^{k_1}z_{hu}\log \pi_{hu}^{(U)},
\end{eqnarray*}
which is directly related to the {\em incomplete log-likelihood} defined
in (\ref{eq:loglik2}) and where $z_{hiv}$ is the indicator function for subject $i$ being in latent class $v$ ($V_{hi}=v$)  and $z_{hu}$ is the indicator function for cluster $h$ being of typology $u$ ($U_h=\xi_u^{(U)}$).
Consequently, $z_{hu}z_{hiv}$ is equal to 1 if both conditions are satisfied and to 0 otherwise.

Usually, $\ell^*(\b\th)$ is much easier to maximize with respect of
$\ell(\b\th)$. The EM algorithm alternates the following two steps until convergence in
$\ell(\b\th)$:
\begin{itemize}
\item \underline{E-step.} It consists of computing the expected value of
the complete
log-likelihood $\ell^*(\b\th)$. In practice, this is equivalent to computing  the posterior expected values of the indicator variables. In particular, we have that
\begin{equation}
\hat{z}_{hiv} = p(V_{hi}=v|\b D) = \sum_{u=1}^{k_U} \widehat{(z_{hu}z_{hiv})}.  \label{eq:pesi1}
\end{equation}
where $\b D$ is a short-hand notation for the observed data. Moreover, we have
\begin{eqnarray*}
\widehat{(z_{hu}z_{hiv})} &=& p(U_h=u,V_{hi}=v|\b D) = p(V_{hi}=v|U_h=\xi_u^{(U)},\b D)p(U_h=\xi_u^{(U)}|\b D)\\ \nonumber
&=&\frac{\pi_{hi,v|u}^{(V)}\prod_{j=1}^J p(y_{hij}|V_{hi}=v)}
{\sum_{v'=1}^{k_V}\pi_{hi,v'|u}^{(V)} \prod_{j=1}^J p(y_{hij}|V_{hi}=v')} \hat{z}_{hu}
\end{eqnarray*}
and
\begin{equation}
\hat{z}_{hu} = p(U_h=\xi_u^{(U)}|\b D) = \frac{\rho_h(\xi_u^{(U)})}{\sum_{h'=1}^H\rho_{h'}(\xi_u^{(U)})},
\label{eq:pesi2}
\end{equation}

\item \underline{M-step.} It consists of updating the model parameters
by maximizing the expected value of $\ell^*(\b\th)$, computed at the 
E-step. More
precisely, while for the individual class weights  an explicit solution exists, for 
the other parameters an explicit solution does not exist.
Therefore, an iterative optimization algorithm of Newton-Raphson type is
used.
The resulting estimates of $\b\th$ are used to update  $\ell^*(\b\th)$
at the next E-step.
\end{itemize}

When the algorithm converges, the last value of $\b\th$, denoted by
$\hat{\b\th}$, corresponds to the maximum of $\ell(\b\th)$ and
then it is taken as the maximum likelihood estimate of this
parameter vector.
It is important to highlight that the number of iterations 
and, in particular, the detection of a global rather than a local
maximum point crucially depend on the initialisation of the EM
algorithm. Therefore, following \cite{bart:07}, we recommend to try
several initialisations of this algorithm. 

The described algorithm 
is implemented in the {\tt R} package {\tt MultiLCIRT}
\citep{bart:bacci:gna}.
We also clarify that  analyses similar to the one here proposed may be performed, alternatively to {\tt MultiLCIRT}, by means of {\tt Latent GOLD} \citep{latentgold}, {\tt mdltm} \citep{mdltm} and  {\tt Mplus} \citep{mplus} softwares, which allow for multidimensional IRT models, discrete latent variables, multilevel data structures, and presence of covariates.

After the parameter estimation, each subject $i$ can be allocated to one of
the $k_V$ latent classes on the basis of the response pattern $\b y_i$ she/he
provided,   her/his covariates $\b x_i$, and the typology of group she/he belongs to. Similarly, each group $h$ can be allocated to one of the $k_U$ latent classes. In both cases, the most common approach is to assign the subject and the group to the class with the highest posterior probability, computed as in equations (\ref{eq:pesi1}) and (\ref{eq:pesi2}), respectively.

\section{Application to the INVALSI dataset}\label{sec:appl}

In this section, we illustrate the application of 
the multilevel finite 
mixture IRT models to the data collected by the two INVALSI Tests. For the purposes of these
analyses, 
the 30 items which assess reading comprehension within the
Italian Test are kept distinct from the 10 items which assess
grammar competency, as the two sections deal with two different
competencies. Therefore, we consider a model with three dimensions: Reading (V1), Grammar (V2), and Mathematics (V3).
Besides, regarding the way of taking the covariate effects into account, we consider
subjects classified according to gender and schools classified according to the geographic area. 
 
In the following, we  first deal with the problem of the model selection, investigating 
the optimal number of latent classes and  the item parameterisation. Then,  we deal with the analysis of the ability 
distribution  and with the assessment of the covariate effects at both levels of the hierarchy.
\subsection{Model selection}\label{sec:selection}
%

In analyzing the INVALSI dataset by the model described in Section
\ref{sec:methods}, a key point is the choice of the number of
latent classes at the individual level and at the group level, that is $k_V$ and $k_U$ respectively. 

In education settings, studentsÕ performances may be classified into one of several categories on the basis of cut scores. The setting of cut scores on standardised tests is a composite judgmental process 
 \citep{Loo:Bou:01, cis:bun:koo:04} whose complexities and nuances are well beyond the scope of this work. For the purposes of the analysis described in the following, it is enough to aknowledge that it is possible to select a different number of groups depending on the adopted judgmental criteria. Here, we adopt a widespread classification of students into three groups (i.e., basic, advanced, and proficient). 

Afterwords, given $k_V=3$ selected at the students' level, we choose the number of school types 
 relying on the main results reported in the literature about finite mixture models \citep{bier:gova:99, mcla:peel:00, fraley:raft:02, nyl:et:al:07}, who suggest the use of the Bayesian Information Criterion (BIC) of \cite{sch:78}. On the basis of this criterion, the selected number of school types is the one corresponding to the minimum value of
\[
BIC = -2\ell(\hat{\b\th})+\log(n)\#{\rm par}.
\]
In practice, we suggest to fit the model for increasing values of $k_U$, under the constraint $k_U\geq k_V$.  
When 
$BIC$  starts to increase, the previous value of $k_U$ is taken as the optimal one. 
Note that, except $k_V$ and $k_U$, the other elements characterizing the model, that is, the item parameterisation and the multidimensional structure of items,  remain fixed.
If one already has some {\em a priori} knowledge about the multidimensional structure of the set of items, then it is convenient to adopt it.
Otherwise, we suggest to select the number of latent classes taking a very general model based on a different dimension for each item. Similarly, we suggest to adopt a basic LC model when you do not have any specific indication about the item parameterisation. For further details about this strategy see, for instance, \cite{bacci:bartolucci:gnaldi:2014}.


In the application described in the present paper, for the selection of the school types (i.e., the number of latent classes at the school level) we fit the LC model with covariates (gender and geographical area) in the three-dimensional version (V1, V2, and V3), in which each item contributes to measure just one ability, 
for values of $k_U$ from 1 to 6. 
The results of this preliminary fitting are
reported in Table \ref{table2}.
On the basis of these results, we choose $k_U = 5$ school types, as in correspondence of this number of classes the smallest BIC value is observed. 

\begin{table}[ht]
\caption{Log-likelihood, number of parameters and BIC values for $k_U = 1,
\ldots, 6$ latent classes for the INVALSI Test; in boldface is the smallest BIC
value.}
\label{table2}
\centering
\begin{tabular}{cccc}
  \hline\hline 
$k$ & $\ell$ & $\#{\rm par}$ & $BIC$ \\ 
  \hline
1	&	-531346.4	&	205	&	1064688	\\
2	&	-530195.9	&	212	&	1062455	\\
3	&	-529947.7	&	219	&	1062027	\\
4	&	-529829.2	&	226	&	1061858	\\
5	&	-529782.3	&	233	&	 {\bf 1061833} 	\\
6	&	-529766.5	&	240	&	1061869	\\
   \hline
\end{tabular}
\vspace*{0.5cm}
\end{table}

After the selection of  $k_V$ and $k_U$ has been made  as described above, alternative models with the given number of classes at the two levels of the hierarchy and the same latent variables have been considered. In particular, we fitted a 1PL model with covariate effects, and a 2PL model with covariate effects (see Table \ref{table3}).

\begin{table}[!ht]
\caption{Model selection: log-likelihood and BIC values for the 1PL model with covariate effects, and the 2PL with covariate effects; in boldface is the smallest BIC
value.}
\label{table3}
\centering
\begin{tabular}{lccc}
  \hline\hline
 model & $\ell$&  $\#{\rm par}$ & BIC \\ 
  \hline
1PL & -533994.6  & 105 & 1069011 \\ 
  2PL & -530039.6 & 169 & {\bf 1061724} \\ 
   \hline
\end{tabular}
\vspace*{0.5cm}
\end{table}

The BIC value of the 2PL model with covariates is smaller than that observed for the 1PL model. Therefore, we retain the 2PL model with $k_V = 3$ and $k_U = 5$, concluding for a better fit of this model in comparison to the 1PL.

\subsection{Distribution of the abilities}  
Here we deal with estimating the parameters of the multilevel 2PL IRT model 
with $k_V = 3$ and three dimensions (V1, V2, and V3) at the students' level, and $k_U = 5$ and one dimension at the school level, that is, the class average weights and the class-specific abilities. 

In the following, abilities are expressed through a standardized scale and class weights are obtained as average values of the estimated individual-specific class weights, denoted by $\hat{\bar{\pi}}^{(V)}_v$ 
and obtained as
\[
\hat{\bar{\pi}}_v^{(V)}=\frac{1}{n}\sum_{h=1}^H\sum_{i=1}^{n_h}\hat{\pi}_{hi,v|u}^{(V)}.
\]
Table~\ref{table4} shows the estimated abilities and average weights for the three classes of students and the three involved dimensions, together with the rank of the abilities in each column (i.e., each test) to facilitate the interpretation. Inspection of these estimates shows that students belonging to class 3 within the two sections (V1 and V2) of the Italian Test and  the Mathematics Test (V3) tend to have the lowest ability level in relation with the involved dimensions. Overall, the weight of low attainment students grouped in class 3 is quite negligible in terms of class proportions, as they count for slightly more than 17 per cent of the students, overall. Besides, students with the highest ability levels belong to class 2, which counts for a little less than forty per cent, while class 1 is a class of examinees with average ability levels over the three involved dimensions. 


\begin{table}[ht]
\caption{Students' level: distribution of the ordered estimated abilities $\hat{\b\xi}_{v}^{(V)}$ for the three involved dimensions  within classes, together with  the average weights ($\hat{\bar{\pi}}^{(V)}$).} 
\label{table4}
\centering
\footnotesize
\begin{tabular}{lccccc}
  \hline\hline
 &  V1&  V2   &V3 &  $\hat{\bar{\pi}}^{(V)}$\\ 
  \hline				
Class 1	&	-0,643	&	1,322	&	1,214	&	0,174	\\
Class 2	&	0,657	&	2,302	&	1,671	&	0,428	\\
Class 3	&	1,988	&	3,564	&	2,214	&	0,398	\\ 
    \hline
\end{tabular}
\end{table}

Besides, we observe that, overall, predicted abilities over the three dimensions tend to be correlated. In fact, the three Spearman correlation coefficients among the three dimensions are always very high (and greater than 0.99) confirming that the three classes tend to group examinees who show consistent levels of ability over the involved dimensions.

At the school level, the distribution of the estimated average abilities $\hat{\xi}_u^{(U)}$ for the five 
chosen types 
(see Table \ref{table5}) allows us to qualify the schools from the worst schools, classified in the Type 1, to the best schools, classified in the Type 5. 
We observe that more than 46.7\% of the schools belong to the best types (Type 4 and 5), whereas only the 16.2\% is classified among the worst ones; the remaining 37.0\% is of  intermediate type.

\begin{table}[ht]
\caption{School level: distribution of the estimated average abilities $\hat{\xi}_u^{(U)}$,  for $u= 1,\ldots,5$, 
and the average weights ($\hat{\bar{\pi}}^{(U)}$).} 
\label{table5}
\centering
\footnotesize
\begin{tabular}{lccccc}
  \hline\hline
 &Type 1	&	Type 2	&	Type 3	&	Type 4	&	Type 5	\\
  \hline				
 $\hat{\xi}_u^{(U)}$ &-1.122	&	-0.353	&	-0.285	&	0.344	&	0.894	\\
$\hat{\bar{\pi}}^{(U)}$& \;0.081	&	\;0.081	&	\;0.370	&	0.351	&	0.116	\\
    \hline
\end{tabular}
\end{table}

\subsection{Effects of level 1 and level 2 covariates} 

As previously stated, in our multilevel setting, covariates express examinees' characteristics (e.g., gender) as well as school characteristics (e.g. geographic area) and therefore we can talk about level 1 and level 2 covariates, respectively. 

In the following, we deal with estimating the regression parameters for the two covariates within latent classes at the students' level and at the school level, to further study the nature of the classes at both levels of the hierarchy and the substantive differences among them. 

Regression parameters for the level 1 covariate (i.e., gender) are estimated  taking as reference class the one characterized by the worst level of estimated average ability over each of the three involved dimensions (Class 1), and category Males as reference category. 
The regression parameters ($\b\zeta_{1uv}^{(V)}$) estimated as in equation \ref{eq:multi1} over Class 2 and Class 3  (0.117 and 0.175, respectively), and the corresponding standard errors (0.053 and 0.057, respectively) show that females tend to be grouped into these classes, and therefore they tend to score higher than Males in the INVALSI Tests. 

Similarly, regression parameters for the level 2 covariate (i.e., geographic area)  are estimated taking Type 1 (i.e. the worst schools)  as a reference class, and category NW (North West) as reference category. For easiness of interpretation, level 2 estimated regression parameters and the corresponding standard errors are not shown here and replaced by estimated probabilities to belong to any of the five types of schools given the geographic area. Results in Table \ref{table6} confirm what preliminary descriptive analysis (see Table~\ref{table1}) already partly revealed, that is, very diverse levels of attainment when taking into account school geographic areas.

\begin{table}[ht]
\caption{Estimated  conditional probabilities ${\hat{\pi}_{hu}^{(U)}}$ to belong to the five types of schools 
given the geographic area.} 
\label{table6}
\centering
\footnotesize
\begin{tabular}{l|ccccc}
  \hline\hline
	 &	Type 1	&	Type 2	&	Type 3	&	Type 4	&	Type 5	\\
  \hline				
NE	&	0.042	&	0.000	&	0.498	&	0.387	&	0.073	\\
NW	&	0.036	&	0.000	&	0.620	&	0.329	&	0.014	\\
Centre	&	0.031	&	0.044	&	0.358	&	0.472	&	0.095	\\
South	&	0.115	&	0.113	&	0.239	&	0.341	&	0.192	\\
Islands	&	0.180	&	0.243	&	0.140	&	0.232	&	0.206	\\
    \hline
\end{tabular}
\end{table}

On the whole, the great majority of the Italian schools tend to be classified into average and high attainment schools. However, schools of the South and the Islands have a relatively higher probability than the schools of the rest of Italy to belong to the best schools (i.e., Type 5). 
Besides, schools of the North West and North East show a very similar profile, as they display high probability to belong to medium attainment schools and high attainment schools (Type 3 and Type 4 schools, respectively). Finally, schools from the South and the Islands have a relatively high probability to be classified among the best schools and, at the same time, to the worst schools (Type 5 and Type 1 schools, respectively). The latter apparently inconsistent result unravels in fact a widespread phenomenon of teachers' cheating in the latter Italian geographic areas. 

\section{Conclusions}\label{sec:conc}

In this article we propose a framework for assessing the relationship between unobserved groups of examinees and schools, and observed characteristics, and establish the ways observed characteristics are related to unobserved groupings, by at the same time accounting for the multilevel structure of our data. 

The data were collected in 2009 by the National Institute for the Evaluation of the Education System
(INVALSI) and refer to two assessment Tests - on Italian language competencies (Reading comprehension, Grammar) and Mathematical
competencies - administered to middle-school students in Italy. 

We refer our analysis to a class of multilevel and multidimensional latent class IRT models obtained as an extension of that  developed by \cite{bart:07}, by accounting for the multilevel structure of the data and the effects of observed covariates at the students' and school levels.
The applied approach has advantages with respect to other approaches which account for observed variables only, on one side, or on just unobserved classes of examinees, on the other side. In fact, at the various levels of the hierarchy, it permits the combined use of information derived from observed group membership (i.e., examinees' gender and school geographic area) and unobserved groupings (i.e., latent classes of examinees and type of school) and, thus, to characterize distinct latent classes of examinees and schools.

Based on this model, we ascertain the existence of latent classes of examinees who show consistent levels of ability over the involved dimensions, and of a few types of schools, from lowest attainment schools to highest attainment ones.
Next, we study the relationship between  observed level 1 and level 2 variables, and latent classes. At the student level, we find that gender has a significant effect on class membership with females who tend to be grouped in the highest attainment groups of students. At the school level, results reveal how and to what extent factors related to school geographic area (i.e., the cheating of Southern teachers) affect the probability for a school to be grouped in a certain school type. 

Overall, the discussed extension of the latent class IRT model developed by \cite{bart:07} to account for the multilevel structure of the data and the covariate effects allowed us to characterise the classes at the two hierarchical levels in such a way that would not have been detectable through other available models.

\vspace{-1.5mm}
\bibliography{biblio}

\bibliographystyle{spbasic}
\end{document}